\documentclass[prb,showpacs]{revtex4}
\usepackage{graphicx}
\newcommand{\BEQ}{\begin{equation}}
\newcommand{\EEQ}{\end{equation}}
\newcommand{\BEA}{\begin{eqnarray}}
\newcommand{\EEA}{\end{eqnarray}}

\newcommand{\p}{\partial}

\newcommand{\nn}{\nonumber }
\newcommand{\Tr}{\mathop{\rm Tr}}
\newcommand{\eps}{{\epsilon}}

\begin{document}
\title{Complexity in Mean-Field Spin-Glass Models: Ising $p$-spin}

\author{A. Crisanti, L. Leuzzi and T. Rizzo}
\affiliation{Dipartimento di Fisica, SMC and INFM, Universit\`{a} di
Roma ``La Sapienza'', 
P.le A. Moro 2, I-00185 Roma, Italy}

\begin{abstract}
  The Complexity  of the Thouless-Anderson-Palmer (TAP)
 solutions of the Ising $p$-spin is investigated in the temperature
 regime where the equilibrium phase is one step Replica Symmetry
 Breaking.  Two solutions of the resulting saddle point equations are
 found. One is supersymmetric (SUSY) and includes the equilibrium
 value of the free energy while the other is non-SUSY. The two
 solutions cross exactly at a value of the free energy where the
 replicon eigenvalue is zero; at low free energy the complexity is
 described by the SUSY solution while at high free energy it is
 described by the non-SUSY solution. In particular the non-SUSY
 solution describes the total number of solutions, like in the
 Sherrington-Kirkpatrick (SK) model. The relevant TAP solutions
 corresponding to the non-SUSY solution share the same feature of the
 corresponding solutions in the SK model, in particular their Hessian
 has a vanishing isolated eigenvalue. The TAP solutions
 corresponding to the SUSY solution, instead, are well separated minima.
\end{abstract} 


\maketitle

\section{Introduction}
Mean-field magnetic models of spin glasses with built-in quenched disorder
display, in general, different kinds of frozen phases selectable by choosing
the type of interaction among their microscopic elements.  The pure phase can
be either ``glass'', described by means of a step function order parameter,
the {\em overlap} $q=q_0\theta(m-x)+q_1\theta(x-m)$, or ``spin-glass'', in
which case the order parameter is a monotonous increasing function $q(x)$.
The overlap $q$ and its conjugated $x$ come out naturally in the replica
approach following the Replica Symmetry Breaking (RSB) Ansatz.\cite{RSB}
 The parameter
$x$ is, then, the replica group index, and $q$ represents the similarity
between replica groups labeled by $x$.  The first phase mentioned is a one
step RSB (1RSB) phase, for which a two levels hierarchy is hypothesized and
replicas can simply lay in the same group ($q=q_1$) or in different groups
($q=q_0<q_1$). The second kind of phase, instead,  is stable only
when the replica symmetry is broken an infinite number of
times, and one refers to it as Full RSB (FRSB) solution.
In a properly said spin-glass, thus,
 any kind of level of similarity can take place, between the extremal values
of $q(x)$, being $q_0$ the minimal and $q_1$ the maximum.
The parameter $q_1$ is called the self-overlap, i.e.
the overlap among replicas in the same group, or Edwards-Anderson\cite{EA}
 parameter.

In the model that we are going to study in the present paper, the Ising
$p$-spin model,\cite{Der,GM,Gard,Rieger,RM} both phases occur. In a cooling
down from high temperature (paramagnetic phase), 
below some point the dynamic quantities become
stuck out of equilibrium, never reaching their static values. 
In this temperature regime the time translational 
invariance of two-time quantities is
lost and aging takes place. 
Cooling further, the high
energy states responsible for the slowing down of the dynamics become more
and more important down to the point where their free energy becomes lower
than the one of the paramagnetic phase. At such temperature
a thermodynamic transition occurs to a 1RSB phase.  
Eventually, at lower temperature, the system
undergoes a second phase transition
to a FRSB spin-glass phase.

Both frozen phases are characterized by a very high number of
stable and metastable states, although of different nature.
  Such a large range of choice is in its turn the consequence of the
disorder and the subsequent frustration characterizing spin-glasses and
causing the onset of many different configurations of spins minimizing the
thermodynamic potential, organized in the configurational space in rather
complicated ways.  In order to describe the structure of the landscape of the
free energy functional a fundamental tool is the so called complexity, else
said, in the framework of structural glasses, configurational entropy.  This
quantity is the quenched average of the logarithm of the number of metastable
states.  In the present work we have performed a thoroughly investigation of
the complexity of the $p$-spin model in the 1RSB phase.

A significant role in the investigation of the properties of the stationary
points of the mean-field free energy landscape is played by a
Becchi-Rouet-Stora-Tyutin fermionic symmetry,\cite{BRST,ZZ} that arises
following a particular formal approach, recalled in Sec. \ref{quattro}. The
initial symmetry of the action involved in the reckoning of the complexity
functional is not always conserved when the complexity saddle point is
evaluated.
In the present work we deepen and continue
 the contents of Ref. [\onlinecite{noiquen}] related to the Ising
$p$-spin.

In Sec. \ref{uno}
we present the Hamiltonian of the model,
we recall its basic properties and we define the complexity
as the Legendre transform of the replica free energy potential.  
In sec. \ref{due} we follow the alternative approach of Thouless, Anderson
and Palmer (TAP) to mean-field disordered models and we complete
the analysis of the TAP complexity already performed by Rieger.\cite{Rieger}
In Sec. \ref{tre} the two qualitatively different complexities
are computed and displayed. There we present a study of the stability
of the states counted by the complexity.
In Sec. \ref{quattro} we give account  for the qualitative differences
between BRST and non-BRST complexities in terms of the spectrum
of the eigenvalues of the Hessian of the TAP free energy.
In Sec. \ref{cinque} we present our conclusions.

\section{Model}
\label{uno}
The Hamiltonian of the model is
\BEQ
{\cal H}=\sum_{i_1<\ldots<i_p}J_{i_1\ldots i_p}\sigma_{i_1}\ldots \sigma_{i_p}
-h \sum_{i=1}^N \sigma_i\EEQ
where the dynamical variables $\sigma_i$ 
are Ising spins and the couplings $J_{i_1\ldots i_p}$ are 
quenched random variables distributed as
\BEQ
P(J_{i_1\ldots i_p})=\frac{N^{p-1}}{\sqrt{\pi p! }}
\exp\left(-\frac{J^2_{i_1\ldots i_p}~N^{p-1}}{p!}
\right)
\EEQ
This is a generalization of the Sherrington-Kirkpatrick (SK) model,
\cite{SKPRL75}
recovered for $p=2$,
to interactions involving more than two spins.
In the following  we set $h=0$, since
the connection between the different
 approaches to the  computation of the complexity
that we will consider here
relies on the absence of external magnetic fields.

In the high temperature regime this model is in a paramagnetic phase.  As the
temperature is decreased, a dynamic transition occurs at $T_d$, with onset of
an aging regime.  Below that temperature two thermodynamic phase transitions
take place. We denote by $T_s$ the critical temperature for the static
transition between the paramagnetic and the intermediate (glassy) frozen phase
and by $T_G$ (Gardner temperature) the one at which the transition to the low
temperature frozen phase (spin-glass) occurs.  The statistical mechanical
properties can be computed applying the replica trick.  The phase displayed in
the temperature range $[T_G,T_s]$ turns out to be a one step Replica Symmetry
Breaking (1RSB) one (even though the replica symmetric solution (RS)
stays stable),
whereas at the Gardner temperature the system reaches a qualitatively
different phase stable only when an infinite number of RSB is performed.
Since, for $p>2$, the paramagnetic, RS phase is stable in
the replica space at any temperature, the transition temperature $T_s$ is
obtained as the one at which the 1RSB free energy becomes lower than the RS
one, that, by the way, in this model
 coincides with the onset of the 1RSB static solution.

In the present paper we will concentrate on the behavior of the
states structure in the 1RSB phase.

\subsection{Intermediate glassy phase in the replica formalism:
complexity as Legendre transform of free energy}

For clearness (and to introduce notation) 
we  very shortly summarize the
basic results already obtained in the seminal paper of Gardner 
\cite{Gard}
and recently used in Ref. [\onlinecite{RM}].
Performing the quenched average over the disordered interaction, 
the free energy is obtained with the replica trick \cite{MPV}
in terms of the overlap
order parameter $q(x)=q_0\theta(m-x)\theta(x)+q_1\theta(x-m)\theta(1-x)$.
In the case of zero external field (inducing $q_0=0$)
it reads:

\BEQ
\beta \Phi=-\frac{\beta^2}{4}\left[1+(p-1)(1-m)~q_1^p-p~q_1^p\right]
-\frac{1}{m}\log\int^\infty_{-\infty}{\cal D}z
\left(2\cosh z\sqrt{\lambda}\right)^m
\EEQ
with $\lambda\equiv \mu~q_1^{p-1}$, $\mu\equiv \beta^2 p/2$ and 
${\cal D}z\equiv dz/\sqrt{2\pi}\exp(-z^2/2)$.
The self-overlap $q_1$ is computed by means of the self-consistency
equation
\BEA
q_1&=&\left<
\tanh^2  z\sqrt{\lambda}
\right>_m
\label{q1}
\\
\left<\left(\ldots\right)\right>_m&\equiv&
\frac{\int^\infty_{-\infty} {\cal D}z
\left(\ldots\right)
\left(\cosh z\sqrt{\lambda}\right)^m}
{\int^\infty_{-\infty}{\cal D}z
\left(\cosh z\sqrt{\lambda}\right)^m}
\EEA
At $T\to T_s$, the Edwards-Anderson parameter $q_1$
jumps discontinuously from $0$ to a finite value
for $p>2$, unlike the SK case
for which the order parameter smoothly grows as the temperature 
crosses the critical value.

Generalizing the definition of entropy in the canonical ensemble
to disordered systems,
the complexity is usually defined as the Legendre Transform of the 
free energy averaged over quenched disorder:\cite{MPRL95}
\BEQ
\Sigma(f)=\beta_e ~f-\beta_e \Phi(\beta_e)
\EEQ
with conjugated variables $f$ and $\beta_e\equiv \beta~m$.
The relationship $f(m)$ (or equivalently $m(f)$) is thus obtained,
at constant temperature, from one of the following formulas:
\BEA
f&=&\frac{\p \beta_e \Phi(\beta_e)}{\p \beta_e}=\frac{\p m~\Phi(m)}{\p m}
\\
\beta_e &=& \frac{\p \Sigma(f)}{\p f}
\label{spPhi}
\EEA
as well as the complexity expression
\BEQ
\Sigma(m)=\beta_e^2\frac{\p \Sigma}{\p \beta_e}
=\beta m^2\frac{\p \Phi(m)}{\p m}
=\frac{(\beta m)^2}{4}(p-1)q_1^p
+\log\int{\cal D}z(2\cosh z \sqrt{\lambda})^m
-m\left<\log 2 \cosh z \sqrt{\lambda}\right>_m
\label{Sigma_1}
\EEQ

\begin{figure}[t!]
\begin{center}
\includegraphics*[width=.49 \textwidth, height=.24 \textheight]{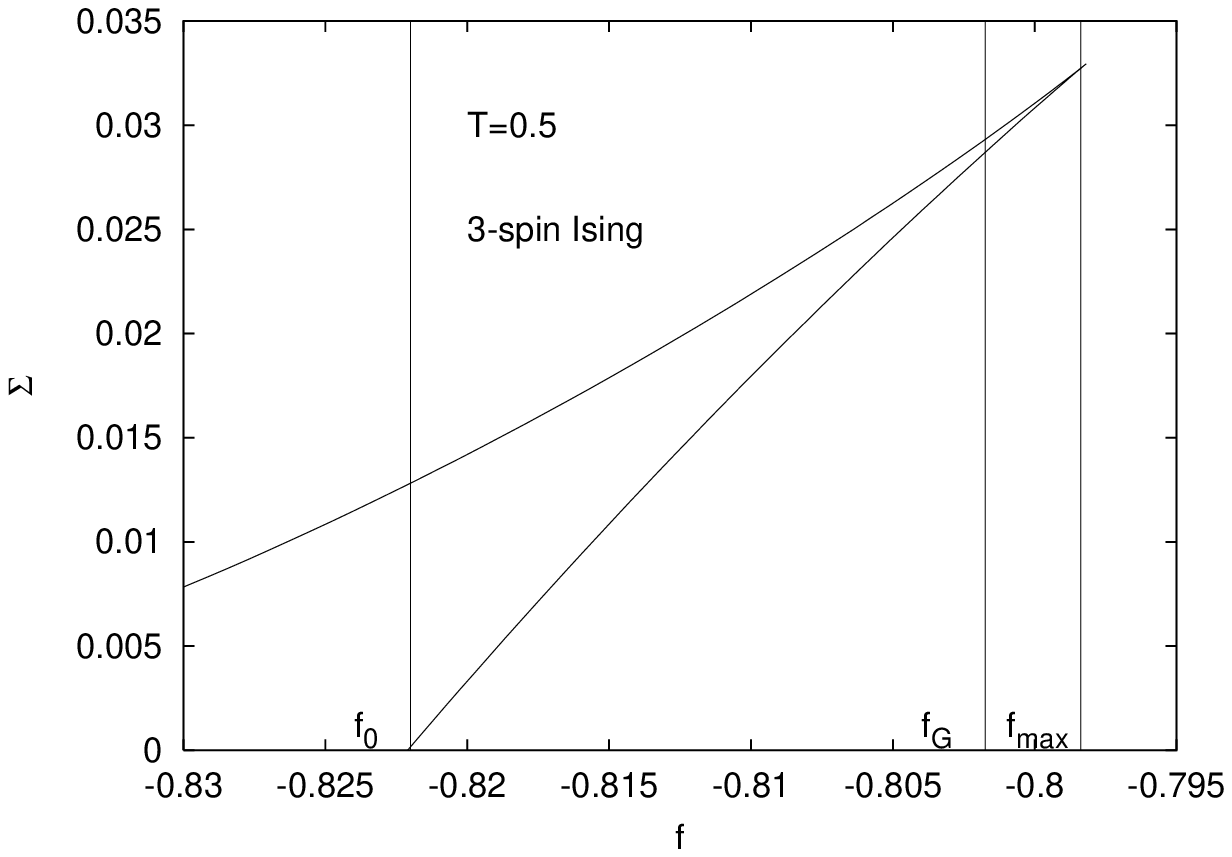}
\includegraphics*[width=.49 \textwidth, height=.24
\textheight]{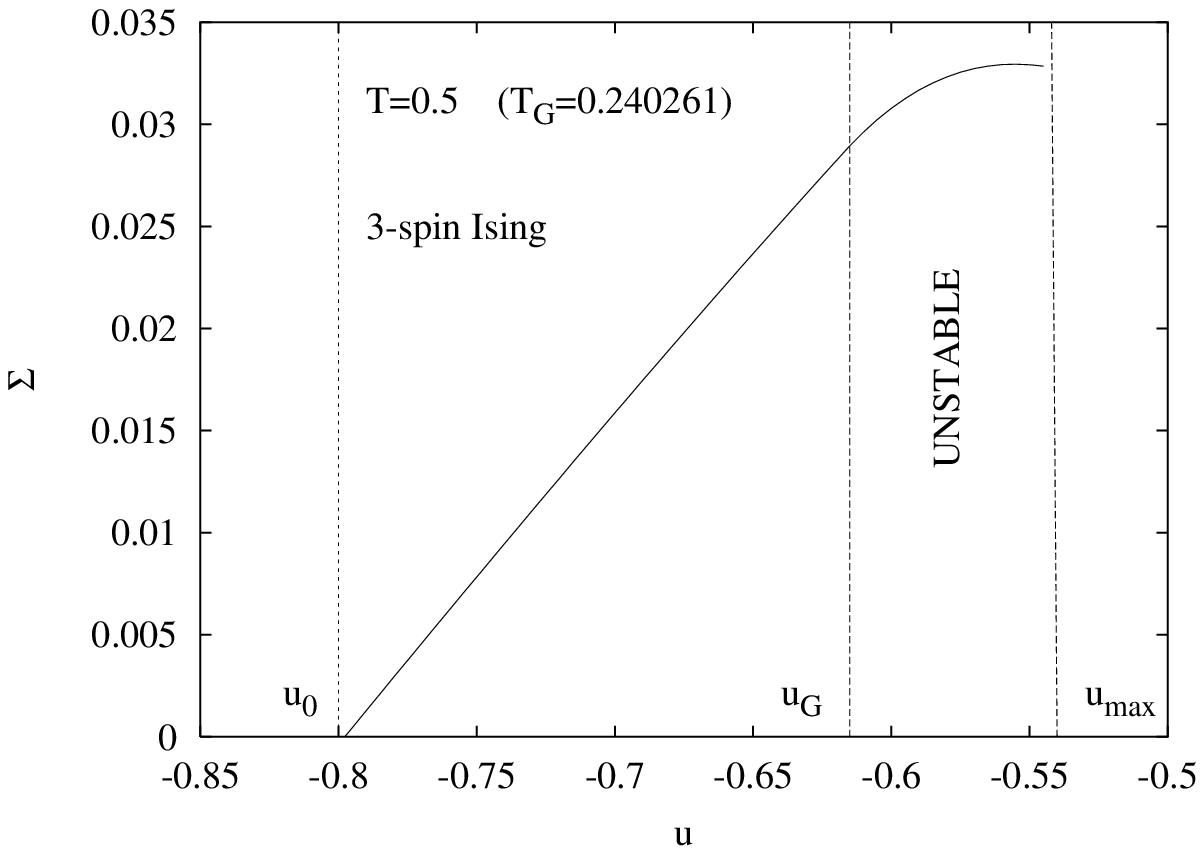}
\caption{The complexity versus $f$ and $u=-m$
computed at $T=0.5$ on the 1RSB static solution
at fixed $m$, Eq. (\ref{Sigma_1}). For $p=3$, $T_s=0.6513$ and $T_G=0.2403$.}
\label{f:S1RSB}
\end{center}
\end{figure}

We notice that the self-consistency equation $\p \Phi/\p m=0$, together with
Eq. (\ref{q1}) yields the equilibrium values of $m$ and $q$,
corresponding with the lowest free energy $f_{\rm eq}(T)$
at which stable states are found at temperature $T$.
Therefore, the complexity vanishes at $f_{\rm eq}$, as lower bound edge.
The higher bound edge, $f_{\rm th}$ (or $m_{\rm th}$),
is instead the point at which its maximum takes place,
 see figure \ref{f:S1RSB}.
Probing the free energy landscape in free energy do not yield,
however, metastable states up to $f_{\rm th}$, as we are now going to recall
(see also Ref. \onlinecite{RM}).

The replicon eigenvalue,
whose positiveness determines the
stability of the phase,\cite{dealme} computed with the 1RSB Ansatz yields
\BEQ
\hat\Lambda=1-\mu(p-1)q_1^{p-2}
\left(1-2q_1+\left<\tanh^4 z\sqrt{\lambda}\right>_m\right)=
1-\mu(p-1)q_1^{p-2}\left<\frac{1}{\cosh^4 z\sqrt{\lambda}}\right>_m>0
\label{replicon}
\EEQ 
The second transition temperature $T_G$ is then determined as the point
at which $\hat \Lambda=0$.  Also, at fixed $T\in[T_G,T_s]$, the replicon
becomes negative above some ``Gardner threshold'' free energy $f_G\leq f_{\rm
  th}$ (figure \ref{f:S1RSB}).  Therefore, the complexity in this temperature
range is well defined only in the interval $[f_{\rm eq},f_G]$.
At $T_s$: $f_G=f_{\rm th}$, at
$T_G$: $f_G=f_{\rm eq}$.

\section{TAP complexity of the Ising $p$-spin model}
\label{due}
{ Another method to derive the complexity, 
inspired to the Boltzmann microscopical interpretation of 
the entropy, is to directly compute the logarithm of   the number
of metastable states
The further step is now, in comparison with ordinary statistical mechanics,
that  the average over the quenched disorder must also be taken.
As a first approximation, to be discussed in the following (see also e.g.
Refs. [\onlinecite{CGPM,noian}]),
we can consider as state a solution of the mean-field equations
for the average site magnetizations, else said Thouless-Anderson-Palmer
(TAP) equations.
}

The computation of the total number of TAP solutions of the Ising
$p$-spin model was performed in Ref. [\onlinecite{Rieger}]; in this section we will
generalize it to the case of solutions of a given free energy.  We
will compute the average number of solutions of the following TAP
equations:

\BEQ
\tanh^{-1} (m_i)={\beta\over (p-1)!}\sum_{j_2,
\dots,j_p}J_{i j_2 \ldots j_p}m_{j_2},
\ldots,m_{j_p}-m_i{\beta^2\over 2}p(p-1)(1-q)q^{p-2}.
\label{tappspin}
\EEQ 

These equations can be obtained considering the behavior of the
cavity field as in Ref. [\onlinecite{TAP}] but also by differentiation of the
following free energy functional:

\BEA
\beta F_{TAP}(\{m\})/N & = & {1\over N}\sum_i\left({1+m_i\over 2}
\ln{1+m_i\over 2}+{1-m_i\over 2}\ln {1-m_i\over 2} \right)+
\nn
\\
& - & {\beta \over N}\sum_{i_1<\ldots<i_p}J_{i_1\ldots i_p}m_{i_1}
\ldots m_{i_p}-
{\beta^2\over 4}((p-1)q^p-pq^{(p-1)}+1)
\label{tapfe}
\EEA

The density of solutions at a given free energy can be expressed as an
integral over the whole $m$-space of a delta function of the TAP
equations:

\begin{eqnarray}\rho(f) & = &\sum_{\alpha=1}^{\cal N} \int \prod_i dm_i\ 
\delta(m_i-m_i^\alpha) \ \delta[F_{TAP}(m^\alpha)-Nf] =
\nn
\\
& = & \int \prod_i dm_i\ \delta(\partial_i F_{TAP}(m)) \ 
|\det (\partial_i \partial_j F_{TAP}(m))| \ \delta[F_{TAP}(m)-Nf] 
\ ,
\label{comp}
\end{eqnarray}
where:
$\p_i$ stays for the partial derivative with respect to $m_i$,
\BEQ
\partial_i F_{TAP}(m))={1\over \beta}g(m_i)-{1\over (p-1)!}
\sum_{j_2,\dots,j_p}J_{ij_2\ldots j_p}m_{j_2},\ldots,m_{j_p}
\EEQ

\BEQ
g(m_i)=\tanh^{-1} (m_i)+m_i{\beta^2\over 2}p(p-1)(1-q)q^{p-2}.
\EEQ

\BEA
\partial_i \partial_j F_{TAP}(m)) & = & \left( {1\over \beta}
{1\over 1-m_i^2} 
+{\beta \over 2}p(p-1)(1-q)q^{p-2}\right)\delta_{ij}-\tilde{J}_{ij}
\nonumber
\\
&+&{\beta\over 2}p (p-1)((p-2)q^{p-3}-(p-1)q^{p-2}){m_i m_j \over 2 N}.
\label{taphess}
\EEA
\BEQ
\tilde{J}_{ij}={1\over (p-2)!}\sum_{k_3,\ldots,k_p}J_{ijk_3\ldots k_p}m_{k_3},
\dots,m_{k_p}
\label{defjtilde}
\EEQ

Notice that the last term in equation (\ref{taphess}) is order
$O(1/N)$, its effect will be discussed below.  The delta function
over the TAP equations $\delta(\partial_i F)$ can be expressed in an
integral form while, after dropping the modulus, the determinant of
the Hessian in Eq. (\ref{comp}) can be expressed in integral form
through a replicated bosonic representation\cite{BMan,Rieger} or
through a fermionic representation.
\cite{PaSou1,PaSou2,K91,CGPM,noian,PR}  The details of { the computation}
in the case of the Ising $p$-spin model can be found
in Ref. [\onlinecite{Rieger}]. 
Here we concentrate only on the last term in
Eq. (\ref{comp}), {\it i.e.} the delta function on the free energy.
Following the original paper of Bray and Moore\cite{BMan} 
we use a non-variational
form of the free energy function (\ref{tapfe}) that is valid only on
the solutions of the TAP equations. This  form is
obtained substituting equation (\ref{tappspin}) in equation
(\ref{tapfe}). We recall that, using this non-variational expression
does not break the supersymmetry of the problem, as noted in
Ref. [\onlinecite{noian}]. 

{{ As an intermediate step we report the expression of $\rho(f)$ in terms of
  a fermionic-bosonic action:
\BEA
&&\rho(f)=\int {\cal D}\psi~ {\cal D}{\overline{\psi}}~ {\cal D}m~{\cal D}x
~e^{{\cal S}(\{\psi,{\overline{\psi}},x,m \})}
\\
&&{\cal S}(\{\psi,{\overline{\psi}},x,m \})=\beta\left\{\sum_{i=1}^N
\left(x_i+\frac{u}{p} m_i\right) \p_i F_{TAP}(m)
+\sum_{ij}{\overline{\psi}}_i\psi_j \p_i\p_j F_{TAP}(m)
+u~F_{TAP}(m)-u~N~f\right\}
\EEA
satisfying the fermionic symmetry:
\BEQ
\delta \psi_i =0\ \ \delta{\overline{\psi}}_i=-\eps x_i\
 \ \delta m_i=\eps \psi_i \ \ \delta x_i =\eps \frac{u}{p} \psi_i
\label{trasf}
\EEQ
where $\eps$ is a fermionic small number.
This invariance, said Becchi-Rouet-Stora-Tyutin (BRST),
 can otherwise be expressed through Ward identities as:
\BEA
&&\left<{\overline{\psi}}_i\psi_i\right>=-\left<m_ix_i\right>+
\frac{u}{p}\left<m_i^2\right>
\label{W1}
\\
&&
u\left<{\overline{\psi}}_i\psi_i\right>=\left<x_i^2\right>-\left<m_ix_i\right>
+\frac{u^2}{p^2}\left<m_i^2\right>
\label{W2}
\EEA
where the average is performed over the measure $e^{\cal S}$.
}}

After some further standard manipulations we
obtain the free energy of a TAP solution as a sum of $N$ local terms
$f(m_i)$ where:

\BEA
\beta f (m) & = & {1\over 2}\ln (1-m^2)-\ln 2+{p-1\over p} m~ \tanh^{-1} (m)+
\nonumber
\\
& - &
{\beta^2\over 4}\left[1+(p-2)q^{p-1}-(p-1)q^p\right]
\EEA

The last expression allows to obtain the generalization of the
equations for the computation of the complexity of
{the Ising $p$-spin model performed by Rieger.\cite{Rieger}}  The
resulting expression is an integral over a finite number of
macroscopic variables that can be evaluated by the saddle point
method. The variational expression of the annealed complexity is:

\BEQ
\Sigma(f)={1\over N}\ln \overline{\rho(f)}= {\rm Ext} -f\,u\,\beta 
- \left( 1 - q \right) \,\left( B + \Delta  \right)  - q\,\lambda  + 
  {\frac{\,{B^2} - {{\Delta }^2} }{2 q^{p-2}\,\left(  p-1 \right) \,\mu }}
+\ln I
\label{refref}
\EEQ
where
\BEQ
I=\int _{-1}^{1}{\frac{1}{{\sqrt{2\,\pi \,\mu \,{q^{p - 1}}}}}}\,
      \left( {\frac{1}{1 - {m^2}}} + B \right) \,
      \exp \left[-{\frac{{{\left( \tanh^{-1}(m) - \Delta \,m \right) }^2}}
           {2\,\mu \,{q^{p - 1}}}} + \lambda \,{m^2} + u\,\beta f(m)\right]\,dm
\label{II}
\EEQ

The saddle point
equations obtained extremizing with respect to $u$,$\lambda$,$\Delta$,$q$ and
$B$ are:

\BEA
f &=&  \langle f(m)\rangle
\label{f}
\\
q & = & \langle m^2 \rangle
\label{q}
\\
\Delta & = & 
  -{\frac{\left( p - 1 \right) \,\mu \,\left( 1 - q \right) \,{q^{p - 2}}}{p}}
+ 
    {\frac{\left( p - 1 \right) \,\langle m\,\tanh^{-1}(m) \rangle }{p\,q}}
\label{Delta}
\\
\lambda & = &
B + \Delta  + {\frac{\left( 2 - p \right) \,{q^{1 - p}}\,
    \left( {B^2} - {{\Delta }^2} \right) }{2\,\left( -1 + p \right) \,\mu }} - 
   {\frac{p - 1}{2\,q}}\,\left( 1 - {\frac{{ \langle{{{\left( m\,\Delta  
- \tanh^{-1}(m) \right) }^2}}\rangle }}{\mu \,{q^{p - 1}}}} \right) +
\nonumber
\\
 &-& 
   {\frac{u\,{{\beta }^2}}{4}}\,\left( p - 2 \right) \,
\left( p - 1 \right) \,{q^{p - 2}} + 
   {\frac{u\,{{\beta }^2}\,p\,\left( p - 1 \right) }{4}}\,{q^{p - 1}}
\label{lambda}
\\
1 - q & =&  
   {\frac{B}{\mu \,\left( p - 1 \right) \,{q^{p - 2}}}}+\left<
 \left( {1\over 1-m^2}+B\right)^{-1}\right>
\label{B}
\EEA
The  brackets $\left<\ldots\right>$
in the previous equations represent averages with
respect to the integrand in Eq. (\ref{II}).

{In order to probe the complexity at different free energy
levels, Eq. (\ref{f}) is relaxed and the remaining equations are solved
at fixed $u$.
Moreover, the request of a positive replicon implies
that, in Eq. (\ref{B}), we choose the solution $B=0$
(see Ref. [\onlinecite{BMan,noian}] for $p=2$ or App. B for any $p$).
}

\section{The Complexity of the Ising $p$-spin model}
\label{tre}

As it was the case in the Sherrington-Kirkpatrick (SK) model,\cite{noian}
also for $p>2$, two different solutions of the above
saddle point equations (\ref{q})-(\ref{lambda}) exist.
In one of them the parameters $q$, $\Delta$ and $\lambda$
satisfy the relations
\begin{eqnarray}
&&\Delta=-\frac{\beta^2}{2}(p-1)u~q^{p-1}
\label{BRST1}
\\
&&\lambda=\frac{\beta^2}{4}\frac{(p-1)^2}{p}u^2q^{p-1}
\label{BRST2}
\end{eqnarray}
deriving from the BRST Ward identities Eqs. (\ref{W1})-(\ref{W2}).
In the second solution, instead, Eqs. (\ref{BRST1})-(\ref{BRST2})
do not hold, i.e. the system is not invariant
under the transformation
Eq. (\ref{trasf}).

We notice that the 
first solution coincides with the static (1RSB) solution, that is stable in
the range of temperature $T\in[T_G(p),T_d(p)]$. Therefore the complexity
coincides with Eq. (\ref{Sigma_1}), as it is clear substituting
relations (\ref{BRST1})-(\ref{BRST2}) into Eqs. (\ref{refref}) and 
(\ref{II}), changing the integration variable in Eq. (\ref{II})
as $m=\tanh(\beta z)$ and eventually  setting $u=-m$.
(see also thoroughly discussion in, e.g., Refs. [\onlinecite{CGPM,noian}]).

\begin{figure}[th!]
\begin{center}
\includegraphics*[width=.49 \textwidth]{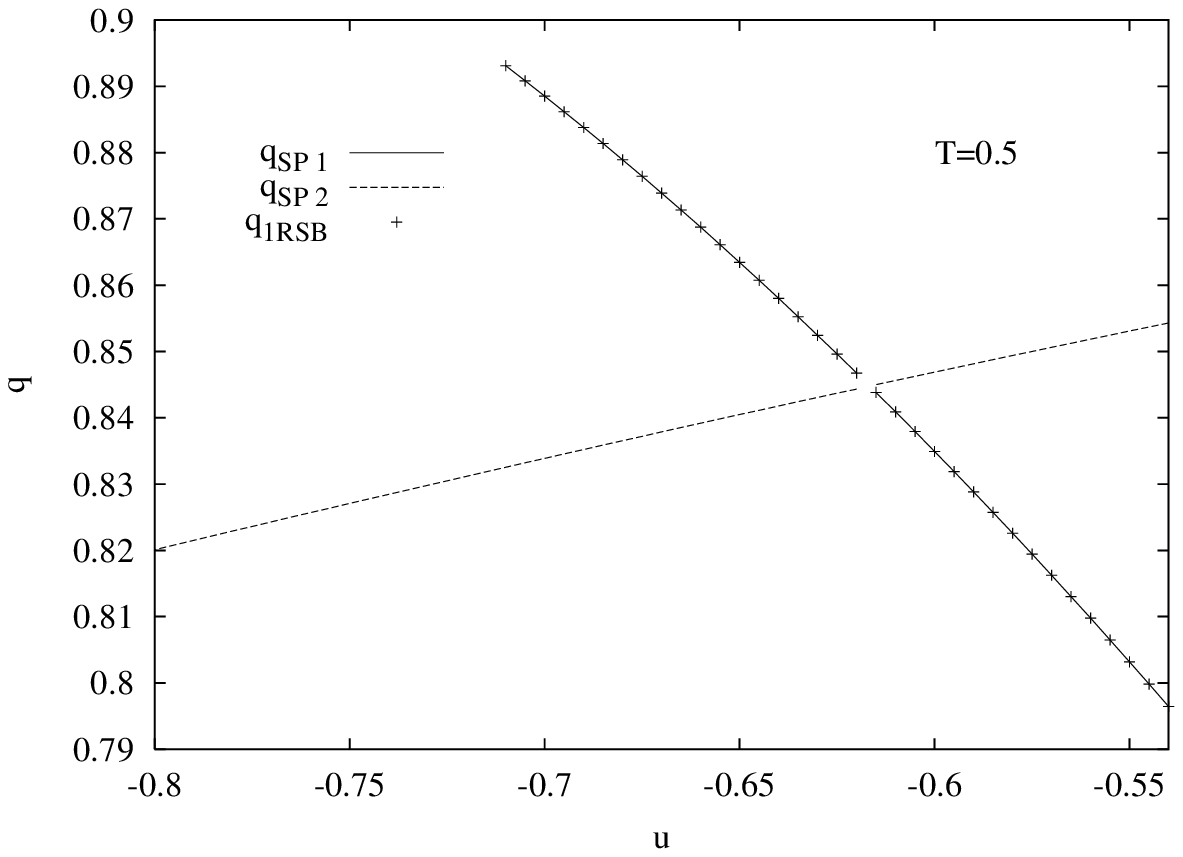}
\includegraphics*[width=.49 \textwidth]{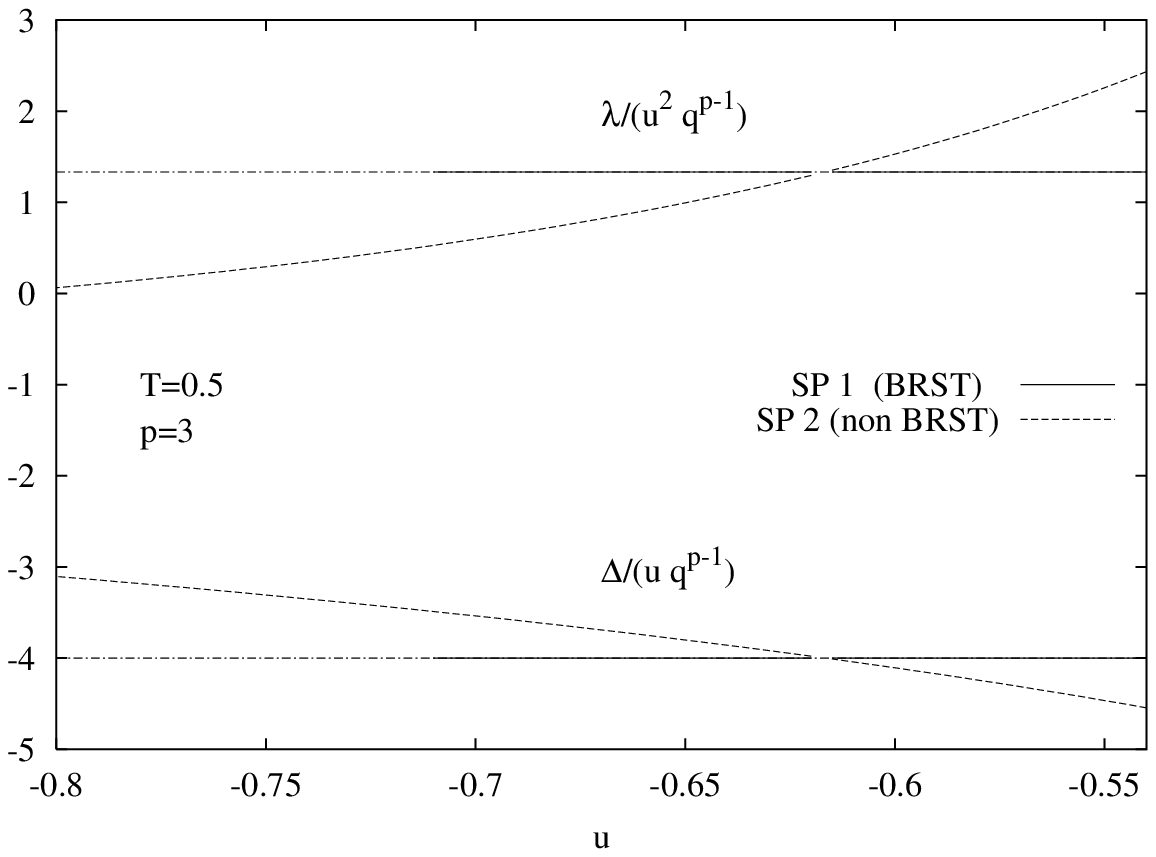}
\caption{The self-overlap $q$ on both complexity saddle points. 
SP1 is BRST symmetric, SP2 is not. The cross points
represent the values obtained in the 1RSB scheme of computation, 
Eq. (\ref{q1}). On the right side the parameters
$\lambda/(u^2 q^{p-1}) $ and $\Delta/(u~q^{p-1})$ are shown versus $u$.
When the BRST relations (\ref{BRST1})-(\ref{BRST2}) are satisfied they are 
constant.}
\label{f:q}
\end{center}
\end{figure}

Setting $p=3$ the first transition temperature is $T_s=0.6513$
and the second one turns out to be $T_G=0.24026$ (obtained equating
 Eqs. (\ref{replicon}) and (\ref{Sigma_1})  to zero). 
In figure \ref{f:q}, for $T=0.5$, we plot the overlap $q$  as a function of $u$
for both solutions of the saddle point equations (\ref{q})-(\ref{lambda})
and for the 1RSB solution.
We also plot the other two parameters
$\lambda$ and $\Delta$ as a function of $q^{p-1}u$ and $q^{p-1}u^2$
respectively. For the BRST solution they are constant 
(Eqs. (\ref{BRST1})-(\ref{BRST2})). 

\begin{figure}[ht!]
\begin{center}
\includegraphics*[width=.49 \textwidth, height=.24 \textheight]{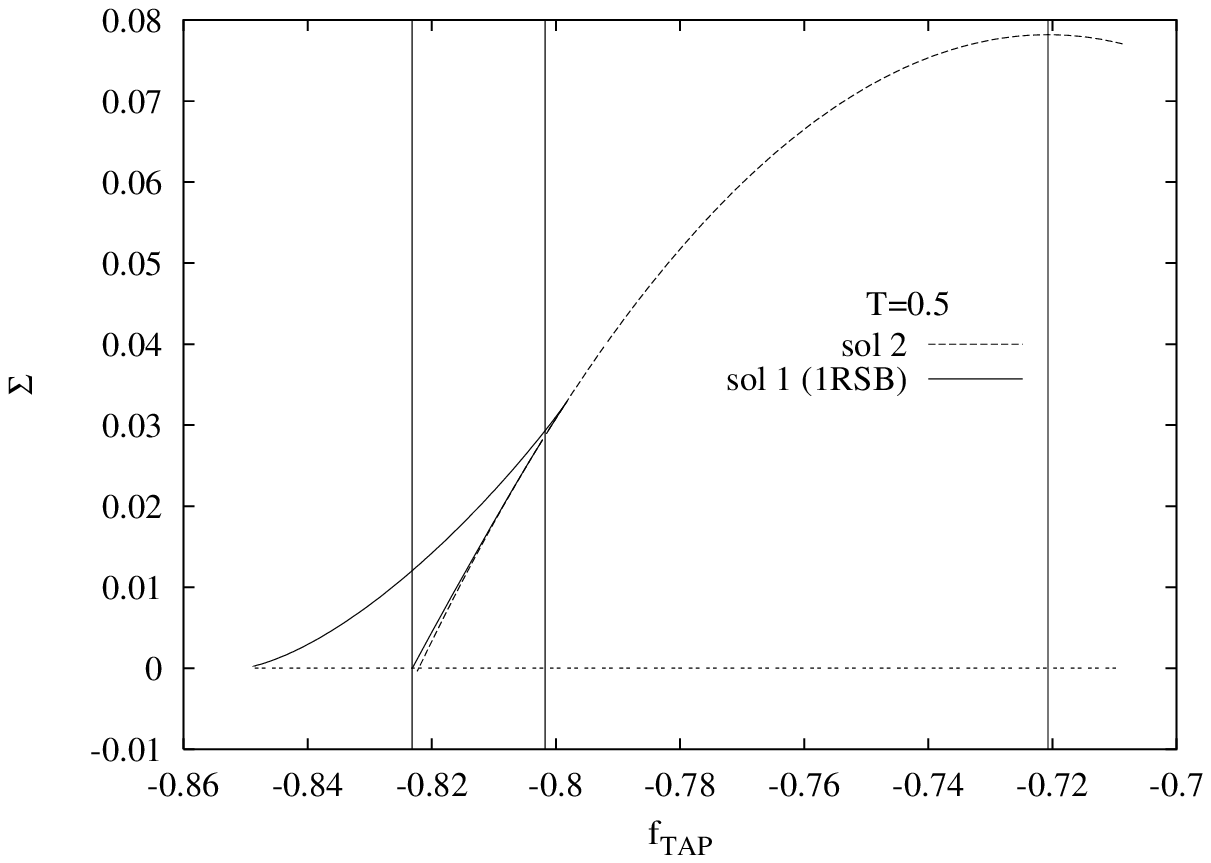}
\includegraphics*[width=.49 \textwidth, height=.24
\textheight]{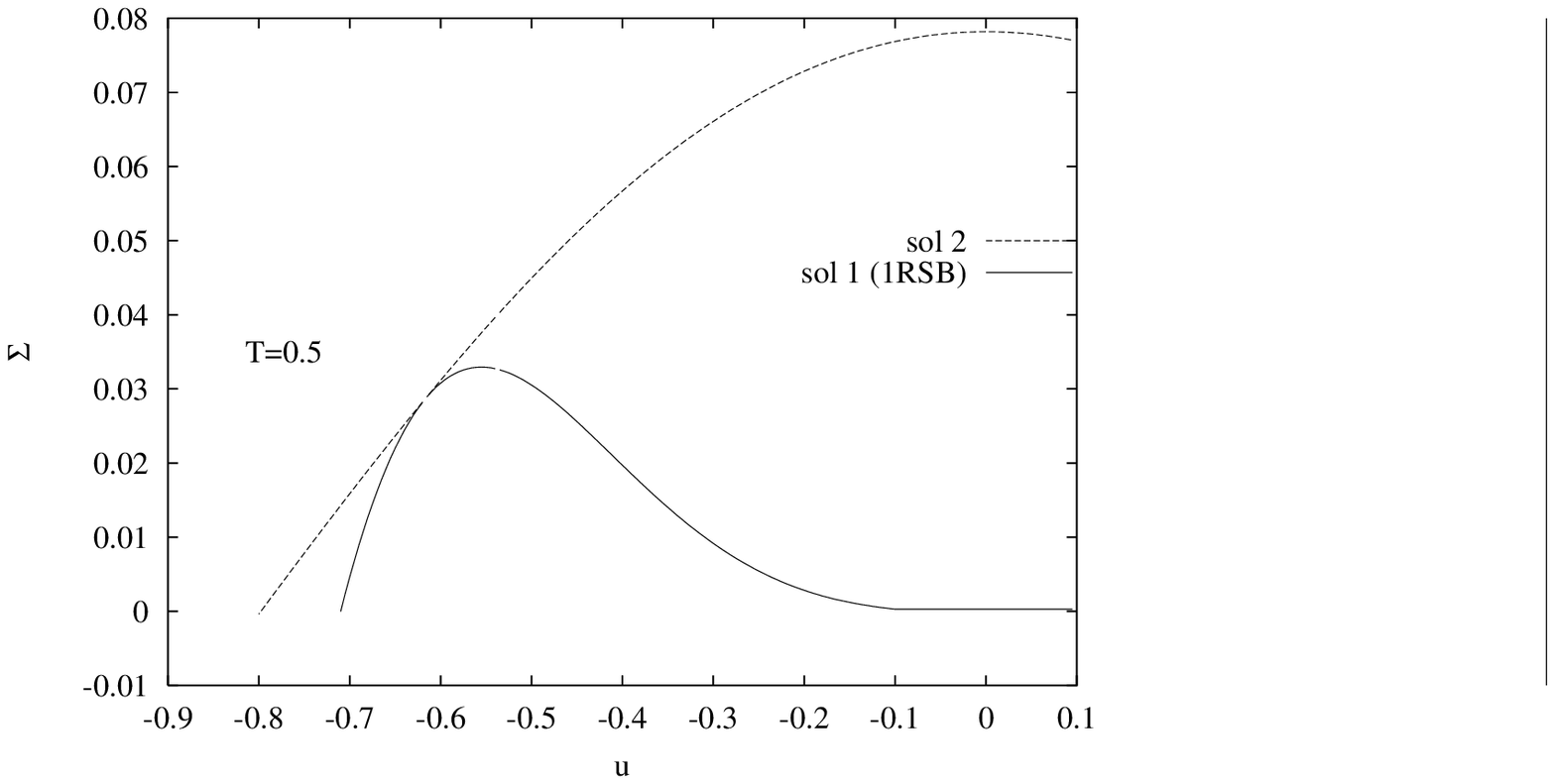}
\caption{The complexity computed as the annealed average of the
logarithm of the number of TAP solutions. Both the BRST and the non-BRST
solutions are plotted. The three vertical lines in the left plot 
stay for, from left to right
respectively, the BRST lower band edge $f_{\rm eq}$, the
crossing point $f_{\rm G}$ and the value of maximum non-BRST complexity. }
\label{f:Sigma}
\end{center}
\end{figure}

In figure \ref{f:Sigma} we plot $1/N\log{\overline{\rho}}$ on both
saddle points.  The behavior in shown both versus $f$ and $u$.  In the $f$
plot the BRST complexity (coinciding with $\Sigma_{1RSB}$) has a cusp. We
stress, however, that the upper, concave,
 branch has to be rejected on physical ground
because there 
\BEQ 
\frac{1}{\beta}\frac{\p f}{\p m}=\frac{\p^2\Sigma}{\p  \beta_e^2}>0 
\EEQ 
and looking at the complexity as a generalization of standard
entropy this corresponds to a thermodynamic instability.
Indeed,
definining the complexity as the Legendre transform of $\Phi$
allows us to look at Eq. (\ref{spPhi}) as the saddle point 
of the integrand of the partition of the states counted with the 
measure $\exp(-\beta_e N f)$, i.e.
\BEQ
 \int df ~\omega(f) ~\exp\{-\beta_e N f\}=
\int df \exp\{-\beta_e N f + N \Sigma(f)\}=
 \int dF \exp\{-\beta_e \Phi(f)\}
\sim \exp\{-\beta_e \Phi(f^{SP})\}
\EEQ
and the condition for the integrand to be a maximum
is
\BEQ
\left.\frac{\p^2}{\p f^2}\Sigma(f)\right|_{sp}=
\frac{\p \beta_e}{\p f} = \beta \frac{\p m}{\p f}  <0 
\EEQ 

At a given temperature the so-computed complexities start
from different lower band edges of  free energy $f_0$.
In the BRST case $f_0=f_{\rm eq}$. In the other case
$f_0<f_{\rm eq}$.
They cross at the value of $u$ (or $-m$ in the replica formalism)
at which the overlap-overlap stability eigenvalue, the {replicon} of Eq.
(\ref{replicon})
becomes negative for the 1RSB solution.
We call this crossing point $u_G$, or $f_G$, the 'Gardner' stability
threshold for low energy metastable states.

\begin{figure}[th!]
\begin{center}
\includegraphics*[width=.49 \textwidth, height=.27 \textheight]
{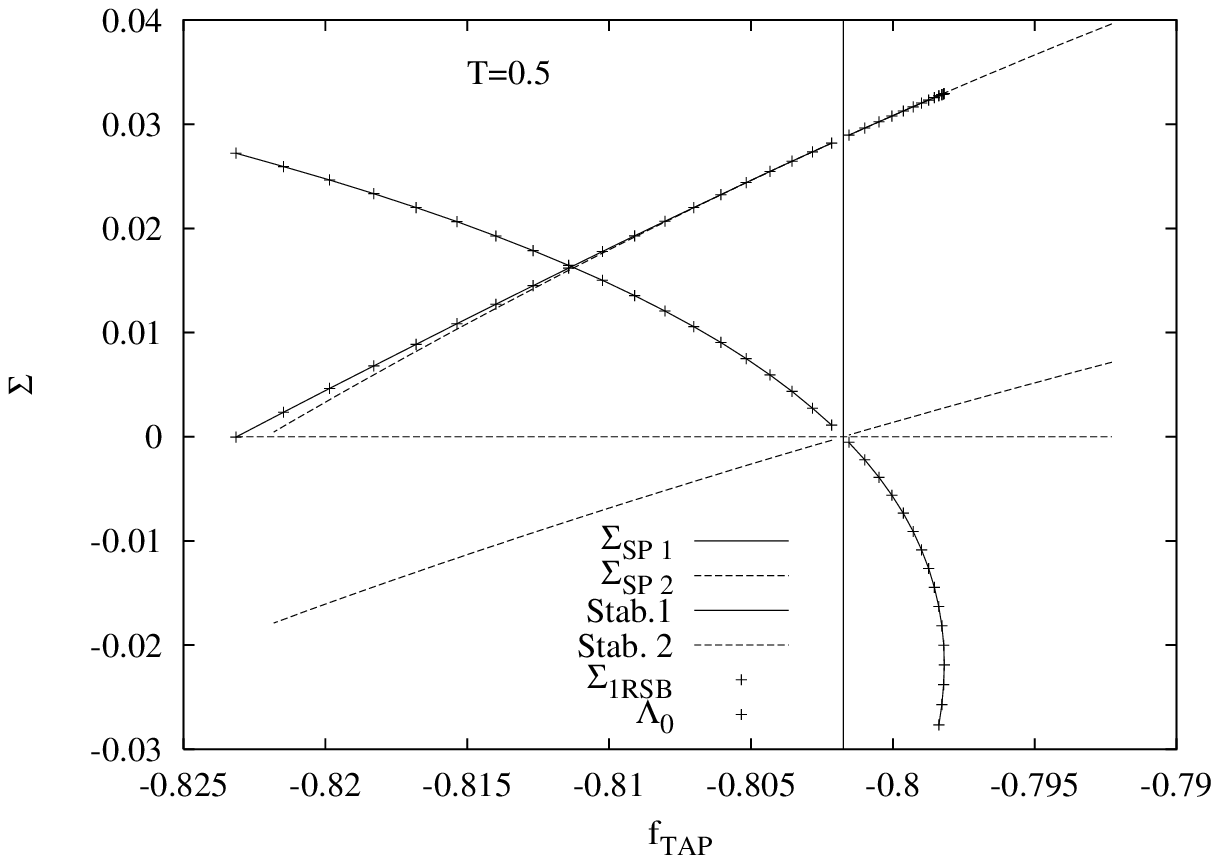}
\includegraphics*[width=.49 \textwidth,height=.27\textheight]
{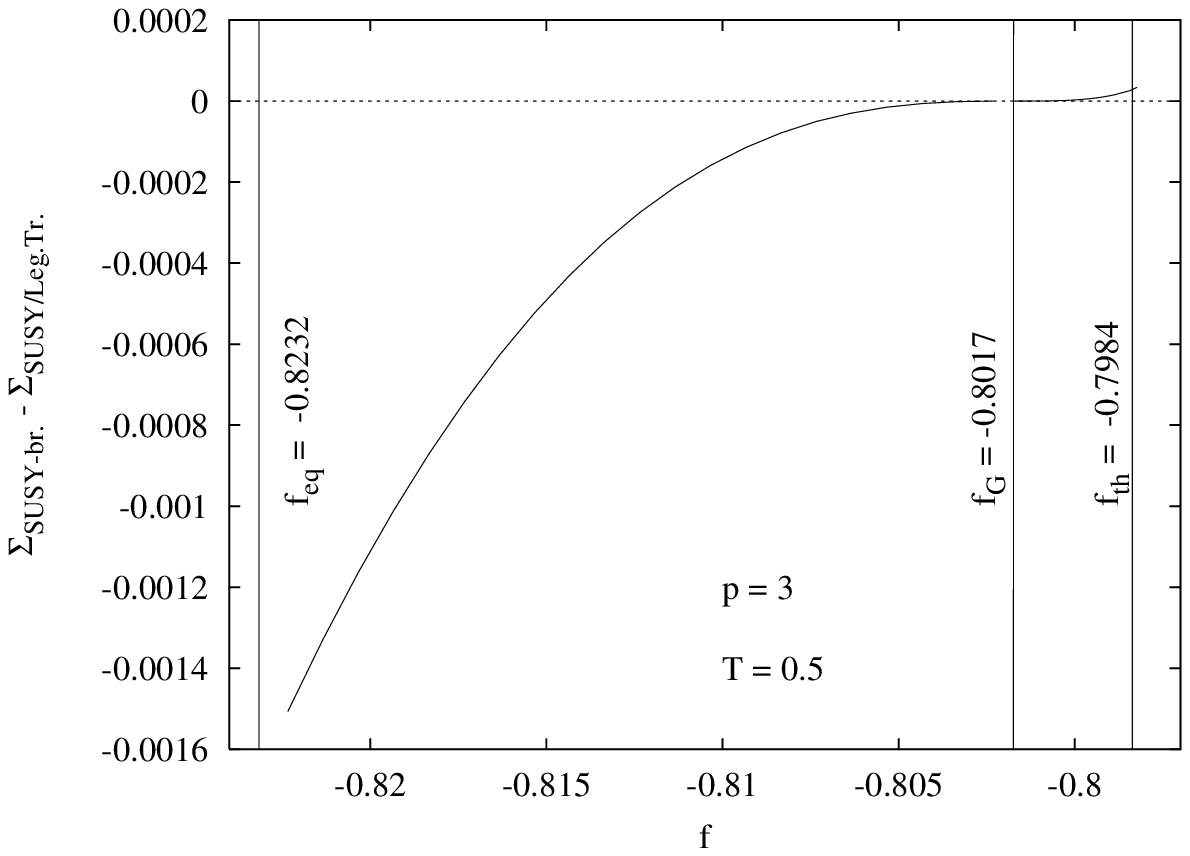}
\caption{
Model with $p=3$. On the left side
the  complexities are plotted at $T=0.5$ vs. $f$.
The continuous line is   for the BRST saddle point of the TAP complexity,
the dashed line for the non-BRST saddle point and the 
the cross points for the static 1RSB complexity (that superimposes on the 
BRST complexity). In the same figure the replicon and  $x_p$ for both
the BRST and the non-BRST complexity saddle points are plotted.
The two $x_p$'s cross in zero at $f=f_{\rm G}$.
On the right side the crossing of the complexities
is expressed more clearly by plotting their difference vs. $f$. 
The three vertical lines represent the stationary 1RSB free energy value
$f_{\rm eq}$, the crossing/1RSB stability value $f_{\rm G}$ and the
threshold value $f_{\rm th}$ at which the cusp of the BRST/1RSB
complexity occurs. 
}
\label{f:cross}
\end{center}
\end{figure}

In terms of TAP equations the stability criterion\cite{Ple1}
 analogous to $\hat \Lambda\geq 0$
is  formulated as the Plefka criterion\cite{Ple1}
\BEQ
x_p=1- \mu (p-1)q^{p-2}\langle (1-m^2)^2 \rangle \geq 0
\label{repli}
\EEQ
A sketch of the derivation is reported is App. A. 
When $x_p$ is computed over the two saddle points one sees that at any $f$
(or $u$) one, and only one, solution is always stable.
The two $x_p$, indeed, cross when they both reach zero, one from
above and one from below, at $f_G$. We explicitely show this in figure
\ref{f:cross} where we also plot the difference in complexity
over the two saddle points.
For $f<f_G$ the BRST complexity is consistent and larger, whereas
for $f>f_G$ the roles are exchanged.

\begin{figure}[ht!]
\begin{center}
\includegraphics*[width=.79 \textwidth, height=.34 \textheight]{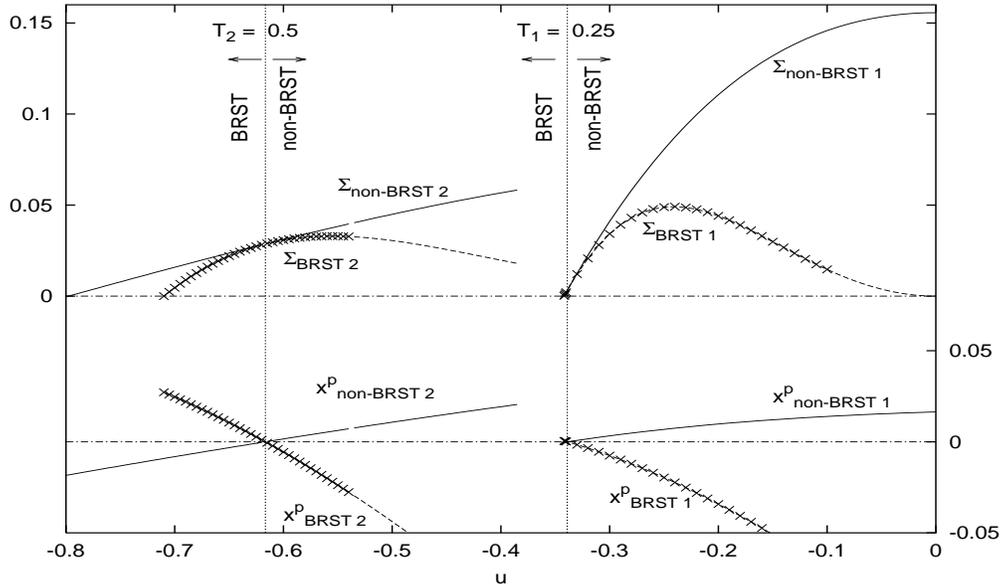}
\caption{The continuous lines
are the complexities and Plefka parameters for BRST and non-BRST
solutions at $T=0.5$ and $T=0.25$ ($T_s=0.6513$,$T_G=0.2403$). 
As the temperature
decreases $f_{\rm G}$ tends to $f_{\rm eq}$. The cross points represent
the 1RSB complexity [Eq. (\ref{Sigma_1})] and the 1RSB replicon 
[Eq. (\ref{replicon})].}
\label{f:2T}
\end{center}
\end{figure}

As the temperature decreases towards $T_G$ the crossing point
shifts towards the lower band edge. To exemplify this,
in figure \ref{f:2T} we have plotted complexities and Plefka
parameter/replicon for the $p=3$ model
both at $T=0.5$ and $T=0.25$.
Exactly at $T_G=0.2403$, $f_G$
and $f_0$ coincide. At this point the BRST complexity,
counting the number of minima\footnote{
By this we mean that the spectrum of the Hessian
(\ref{taphess}) is strictly positive, with no zero eigenvalue.
Different is the case of the ``marginal states'' counted
by the non-BRST complexity that  display a zero eigenvalue.
}
 of the free energy landscape turns out to be physically 
inconsistent on all states but those at $f=f_{\rm eq}$.
Lowering further the temperature we end up in  a FRSB spin glass phase,
for which both the 1RSB solution and the annealed complexity 
are ill defined. In this region the system behaves exactly
as the SK model. This means that, even performing the proper FRSB 
quenched  average, the BRST complexity has to be rejected
for any value of $f$ apart from $f_{\rm eq}$ (computed in the FRSB phase).
\cite{CLPR3}
The only likely candidate as a metastable states counter
remains the non-BRST complexity.

\subsection{Deepening on The Crossing Point}

In appendices A and B  we
recall that the condition of positivity of $x_P$ is recovered also in
the present context as a condition of physical consistency of the TAP
solutions and of analytical consistency of the $B=0$ solution.  Thus
the identification of the crossing point with the point where the
replicon vanishes is crucial because it ensures that at all free
energies we can choose a solution with a positive value of the
replicon. This identification can be justified analytically.  The
solution at a given value of $f$ can be continued in a unique way if
the Hessian of (\ref{refref}) with respect to $B$, $\lambda$,
$\Delta$,$q$ and $u$ has no zero eigenvalue, hence a necessary
condition to locate a crossing point is that the Hessian must have at
least a zero eigenvalue.  In the following we will show that the
vanishing of the replicon or equivalently of the parameter $x_P$ is a
sufficient condition to have a zero eigenvalue in the Hessian.  We
start from the following important relationship \cite{BMan} that can
be derived from the saddle point equation:

\BEQ
{\partial \Sigma\over \partial B}+{\partial \Sigma \over \partial \lambda}
=B\left({\frac{1}{\mu \,\left( p - 1 \right) \,{q^{p - 2}}}}
-\langle {(1-m^2)^2\over 1+B (1-m^2)}\rangle\right)
\EEQ

It is easily seen that deriving the previous equation with respect to
the five parameter $\lambda$, $\Delta$,$q$, $u$, $B$ and then setting
$B=0$ we have:

\BEA
{\partial^2 \Sigma\over \partial B \partial \lambda}
+{\partial^2 \Sigma \over \partial \lambda^2} & = & 0
\\
 {\partial^2 \Sigma\over \partial B \partial \Delta}
+{\partial^2 \Sigma \over \partial \lambda \partial \Delta} &  = & 0
\\
  {\partial^2 \Sigma\over \partial B \partial q}
+{\partial^2 \Sigma \over \partial \lambda \partial q} &  = & 0
\\
  {\partial^2 \Sigma\over \partial B \partial u}
+{\partial^2 \Sigma \over \partial \lambda \partial u} &  = & 0
\\
  {\partial^2 \Sigma\over \partial B^2}
+{\partial^2 \Sigma \over \partial \lambda \partial B} &  = & 
\left({\frac{1}{\mu \,\left( p - 1 \right) \,{q^{p - 2}}}}
-\langle {(1-m^2)^2}\rangle\right)
\EEA

Notice that the r.h.s. of the last equation is proportional to the
replicon Eq. (\ref{repli}), therefore if the replicon vanishes the
r.h.s.'s of the last five equations are all zero, meaning that two
columns of the Hessian of (\ref{refref}) sum up to zero and therefore
it must have at least a zero eigenvalue.

\section{Supersymmetry and the Spectrum of the Hessian}
\label{quattro}
In this section we will discuss the TAP complexity of the Ising
$p$-spin model within the supersymmetryc (SUSY) framework. 
 The main
result is that 
 {\it much as in the SK model , the TAP solutions
described by the non-SUSY solution valid at high free energies display
a vanishing isolated eigenvalue in the spectrum of the Hessian}.

This result, originally obtained for the SK model in
[\onlinecite{ABM}], was proven rigorously in [\onlinecite{PR}] where
it was shown that it is a consequence of SUSY breaking.
 As such the
extension to a SUSY-breaking solution in a different model like the
one we are considering here is straightforward. The rest of this
section is rather technical and the reader not interested in the
details can safely skip to the conclusion.
 
 The details of the
calculations are similar to those performed for the SK model
\cite{K91,CGPM,noian,PR,R} and will not be presented.  In particular,
a self-contained derivation of the final results would be a trivial
rewriting of the results of [\onlinecite{PR}] and [\onlinecite{R}],
therefore we will simply recall them and discuss the basic steps of
their generalization to the Ising $p$-spin model.  The starting point
is the expression of the determinant in (\ref{comp}) by means of a
fermionic representation. Then the complexity is expressed as an
integral over an action depending on $4N$ fermionic and bosonic
variables .  The action is invariant under the so called
Becchi-Rouet-Stora-Tyutin (BRST) transformation \cite{BRST,ZZ}. This
transformation mixes bosonic and fermionic variables {\it i.e.} it is
a SUSY transformation.  Then, through odd and even
Hubbard-Stratonovich transformation the complexity is expressed as an
integral over the exponential of a macroscopic action depending on
four bosonic and four fermionic macroscopic variables $\{r,t
,\lambda,q, \mu, \overline{\mu},\rho, \overline{\rho} \}$.  The
macroscopic action posses a SUSY as well \cite{K91,PR}.  Setting the
fermionic variables to zero and making proper changes of variables one
recovers the expression (\ref{refref}) for the Complexity.
 
 The
integral over the action can be evaluated by the saddle point
method.
 As we have shown in the previous sections the resulting
action admits two solutions, much as the one of the SK model. One of
these solutions is SUSY while the other is not but at variance with
the SK model none of them is good in the whole range of free
energies. Indeed the SUSY solution is correct at low free energies,
particularly at the equilibrium value, while the non-SUSY solution is
correct at high free energy and in particular must be considered to
describe the total complexity.  
 Setting $B=0$ the SUSY
relationships expressed in the variables of (\ref{refref}) are 

\BEQ
\Delta = -{\frac{u\,\mu \,\left( p - 1 \right) \,{q^{p - 1}}}{p}}\ \
;\ \ \lambda = {\frac{{u^2}\,\mu \,{{\left( p - 1 \right) }^2}\,{q^{p
- 1}}}{2\,{p^2}}}
\label{eqsus}
\EEQ

Note that according to these equation the BRST relationships at $u=0$
are $\Delta=0$ and $\lambda=0$, and imply a zero total complexity, the
non-SUSY solution instead predicts a finite complexity while the
parameter $\Delta$ and $\lambda$ satisfies some proper SUSY Ward
identities \cite{R} different from (\ref{eqsus}).
All the properties of the non-SUSY solution derived from SUSY
violation in the SK model apply as well in the case of the Ising
$p$-spin model.  
 In particular it is straightforwward to generalize
the arguments of [\onlinecite{K91}] and [\onlinecite{PR}] showing that
the expansion in power of $1/N$ of the prefactor to the exponential
contribution vanishes at all order at all free energies.
 In the
following we concentrate on the TAP spectrum. The computation of the
resolvent of the Hessian (\ref{taphess}) without the last $O(1/N)$
term can be made following the same lines of [\onlinecite{Ple2}]. 

The discussion greatly simplifies noticing that the term
$\tilde{J}_{ij}$ in (\ref{taphess}) is a random term much as $J_{ij}$
in the SK model but with a variance dependent on the
self-overlap. This effect can be simply accounted for through a
rescaling of the temperature. Thus we obtain the same picture of the
SK model \cite{Ple2}: {\it the extensive part of the spectrum is {\it
always} positive at any value of the magnetizations and is zero only
if the following parameter is zero}

\BEQ
x_p=1- \mu (p-1)q^{p-2}  {1\over N}\sum_i(1-m_i^2)^2 .
\EEQ

The parameter $x_P$ controls the value of the susceptibility of the TAP
equations, only if $x_P$ is greater or equal than zero the
susceptibility of the TAP solutions has the physical value
$\chi=\beta(1-q)$, see also discussion in appendix A.  In this context the 
condition $x_P \geq 0$ appears
only as a condition of physical consistence for the TAP solution,
however it can be shown that it is also a condition of analytical
consistence for the computation of the complexity once the solution
with $B=0$ is adopted in (\ref{refref}), see appendix B and the discussion in
[\onlinecite{BMan,noian,noiquen,Ple2}]. It can be also derived as a stability
condition in the replica framework \cite{Gard}.

As originally noted in [\onlinecite{ABM}] in the context of the SK
model, the last $O(1/N)$ term in the Hessian (\ref{taphess}) splits
the lowest eigenvalue outside the continuous band of a finite amount
\cite{ABM,PR,R}. This property is related to the fact that this term
is proportional to the projector $P_{ij}=m_i m_j$. The effect of the
isolated eigenvalue on the determinant of the Hessian depends on
proper bilinear forms.  In particular if the quantity \BEQ {1 \over N}
\sum_{ij} m_i \left( {\partial^2F \over \partial m \partial m}
\right)^{-1}_{ij}m_j
\label{hessdive}
\EEQ 
is divergent then the isolated eigenvalue must be zero. In [\onlinecite{PR}]
it was shown that this result holds for the BM solution in the SK
model because of the SUSY violation and the argument can be easily applied 
in the present context.
First of all we notice that the effect of a projection term on a
matrix  does not depends on the
model considered, therefore we arrive at the same conclusion that if
the quantity (\ref{hessdive}) is divergent the isolated eigenvalue is
zero. This quantity can be expressed as the
average of microscopic fermionic variables and then as the average of
macroscopic fermionic variables:
\BEQ
{1 \over N} \sum_{ij} m_i \left( {\partial^2F \over \partial m \partial m}
\right)^{-1}_{ij}m_j \propto {1 \over N} \sum_{ij} \langle m_i
\overline{\psi}_i \psi_j m_j \rangle \propto \langle \overline{\mu}\rho
\rangle
\EEQ
The final result is obtained noticing that  this average is divergent because
at the denominator the
integral over the fermionic variables produces a zero prefactor to the
bosonic exponential contribution while at the numerator the prefactor
is finite because of the presence of the fermionic variables on which
the average is performed, see eq. (58-60) in [\onlinecite{PR}].

Finally let us comment on an interesting aspect that is specific of the
Ising $p$-spin model. In [\onlinecite{R}] it was shown that while the relevant
TAP solutions have a vanishing isolated eigenvalue nevertheless there
is an exponential number of couples of TAP solutions with a well
defined value of it. These couples of states can be continued upon a
variation of the external parameters, {\it e.g.} the temperature, and
as a consequence {\it the complexity is continuous}.
According to  the work of Rieger, however, the complexity (non-BRST)
of the Ising $p$-spin model  presents a discontinuity at
a given  temperature,  $T_c(p)$.
This can be accounted if the Hessian
of (\ref{refref}) has a singular behavior at $T_c(p)$
as the figures 1 and 2 of Ref. \onlinecite{Rieger} suggest.
As a
consequence the bell shaped curve $\Sigma(q)$ of [\onlinecite{R}] collapses
to a delta function, meaning that $T_c(p)$ is the only temperature
where {\it all} the TAP solutions have a zero isolated eigenvalue.

\section{Conclusion}
\label{cinque}
In conclusion, we have investigated the complexity of the TAP solutions of the $p$-spin model
as a function of their free energy in the range of temperatures where the static solution is 1RSB.
At any free energy we have found two solutions of the saddle point equations, one
 is non-BRST, while the other
satisfies the BRST symmetry and it is the one computed in the
1RSB framework in \onlinecite{RM}.
At each free energy we can select the correct solution considering the stability of the replicon, indeed where one is solution is stable the other is not.
The two solutions coincide at the value of the free energy $f_G$ where the replicon exaclty vanishes.
Above $f_G$ the non-BRST solution must be chosen while below $f_G$ the BRST one holds.
This implies that below $f_G$ the states computed by
the complexity are proper minima, with all positive
eigenvalues, whereas above it the large majority of 'states'
(exponentially large with $N$) are stationary points of the TAP
landscape with a zero eigenvalue in the TAP Hessian, possibly implying
a qualitatively different dynamic behavior of the system evolving
above and below the Gardner threshold.  We also recall that in
[\onlinecite{R}] it was shown the in the non-SUSY phase there are also
well-defined minima and saddles of order one.  The number of these TAP
solutions with a definitively non-zero isolated eigenvalue is
exponential in $N$ although with a lower complexity in comparison with
the one of the marginal states.
We stress that, unlike in the SK
model, in the present case the annealed complexity is exact, yielding
the right quantitative values at any $f$ and that the stability
condition holds in the whole dominion.  We notice that in a simpler
model (without Gardner threshold, nor lower lying FRSB frozen phase)
i.e. the spherical $3$-spin model perturbed by a four spins
interaction, quite recently presented in Ref. [\onlinecite{cav}], a
similar phenomenon occur.  Also there the BRST and the non-BRST
complexities cross where the Plefka parameter related to each of them
vanishes.

In Ref. \onlinecite{RM}, where $f_G$ was introduced, it was
conjectured that above the Gardner threshold an exponential number of
states continues to exist, their organization being of the FRSB type.
What we have found here is rather a SUSY/non-SUSY transition such
that, at high free energy ($[f_G,f_{\rm th}]$), the majority of
'states' shares the property of having an isolated zero eigenvalue and
this feature is connected to the violation of the BRST relations.  In
the SK model the only likely complexity for the whole free energy
domain in the whole temperature range below the critical point is
non-BRST (this is also the case for the Ising $p$-spin model below
$T_G$).  The only difference with the case here presented being that
the annealed complexity is just an approximation for FRSB systems and,
at least for the lowest values of $f$, the complexity must be
corrected by means of a quenched computation.  Indeed, in agreement
with Montanari and Ricci-Tersenghi\cite{RM}, some kind of states are
present even above $f_G$ and their number exponentially grows with the
size of the system.  However, according to our results, no actual
phase transition to a FRSB phase occurs but a different organization
of the ``valleys'' and ``passes'' of the free energy landscape at high
altitude.
However the problem of the cluster complexity as suggested in [\onlinecite{RM}] remains open.
The SUSY/non-SUSY transition that we have found here may occur also in models
defined on locally tree-like lattices like those considered in
optimization problems. A theory for the computation of the
complexity in tree-like models has been recently presented in
[\onlinecite{R2}] through the cavity method; in that context the
transition corresponds to the development of non-zero $z$-fields.

\acknowledgments
We thank Andrea Cavagna, Andrea Montanari and Federico Ricci-Tersenghi
  for useful discussions and  for a careful reading of the manuscript.

\section*{Appendix A: Stability analysis}
\label{appA}
The stability is governed by the Hessian Eq. (\ref{taphess}), i.e. the inverse
susceptibility matrix
\BEA
\chi_{ij}^{-1}&=&\frac{h_i}{m_j}=
\partial_i \partial_j F_{TAP}(m)) =  \left( {1\over \beta}
{1\over 1-m_i^2}
+{\beta \over 2}p(p-1)(1-q)q^{p-2}\right)\delta_{ij}-\tilde{J}_{ij}
\\
\tilde{J}_{ij}&=&
{1\over (p-2)!}\sum_{k_3,\ldots,k_p}J_{ijk_3\ldots k_p}m_{k_3},
\dots,m_{k_p}
\EEA
To perform the stability check following the lines
of Refs. [\onlinecite{Ple1,Ple2}]
we notice that
\BEA
{\overline{\tilde{J}_{ij}^2}}&=&
\frac{1}{((p-2)!)^2}{\overline{\left(
\sum_{k_3,\ldots,k_p}J_{ijk_3\ldots k_p}m_{k_3},\dots,m_{k_p}
\right)^2}}
\\
\nn
&=&\frac{1}{((p-2)!)^2}(p-2)!
\sum_{k_3,\ldots,k_p}
{\overline{J^2_{ijk_3\ldots k_p}}}
m^2_{k_3},\dots,m^2_{k_p}
+O\left(\frac{1}{N^2}\right)
\\
\nn
&=&
\frac{1}{(p-2)!}N^{p-2} \frac{p!}{2 N^{p-1}} q^{p-2}
+O\left(\frac{1}{N^2}\right)
=\frac{p(p-1)q^{p-2}}{2}\frac{1}{N}+O\left(\frac{1}{N^2}\right)
\EEA

Defining $\alpha_p\equiv p(p-1)q^{p-2}/2$ and
\BEA
K_{ij}&\equiv&\frac{\tilde{J_{ij}}}{\alpha_p}
\\
\tilde{\chi}^{-1}_{ij}&\equiv& \frac{\chi_{ij}^{-1}}{\alpha_p}
=\delta_{ij}\left(
\frac{1}{\alpha_p}
\frac{1}{\beta(1-m_i^2)}
+\alpha_p \beta (1-q)\right)
-K_{ij}
\EEA
and introducing the resolvent
\BEQ
\tilde{R}(z)=\frac{1}{N}\Tr\left(
z-\mathbf{\tilde{\chi}}^{-1}
\right)
\EEQ
satisfying the property
$\Im \tilde{R}(z)>0$ for $\Im z<0$, the analysis of Refs. 
[\onlinecite{Ple1,Ple2}] yields that the physical solution is given by
\BEQ
\tilde{R}(z)=z\left(1-\frac{1}{x_p}\right)-\alpha_p\beta(1-q)
\EEQ
with the condition
\BEQ
x_p=1-\frac{\alpha_p^2}{N}\sum_{i=1}^N (1-m_i^2)^2>0
\EEQ




This way the local suceptibility takes the right form
\BEQ
\chi_l\equiv \frac{1}{N}\Tr \mathbf{\chi}=\frac{1}{\alpha_p N}
\Tr\mathbf{\tilde{\chi}}=\frac{1}{\alpha_p}
\left.\left(-\Re(\tilde{R}(z))\right)\right|_{z=0}=\beta(1-q)
\EEQ

The condition $x_p>0$ over the $\{m\}$ configurations
is equivalent to the replicon positivity in the framework of replicas. 

\section*{Appendix B: 
Connection between the $B=0$ complexity saddle point and 
the Plefka parameter.}
\label{appB}

Performing the computation of the determinat of the 
Hessian, Eq. (\ref{taphess}),
one obtains
\BEQ
\det \chi^{-1}_{ij}=
\int\prod_{i=1}^N d\eta_i~d{\overline{\eta}}_i
\exp\left\{
-\sum_{i<j}\tilde{J}_{ij}
\left(
{\overline{\eta}}_i\eta_j+ {\overline{\eta}}_j\eta_i
\right)
+\sum_{i=1}^N\left(
\frac{1}{\beta(1-m_i^2)}+\alpha_p^2\beta(1-q)
\right)
\right\}
\EEQ
Carrying out the average over the effective couplings
$\tilde{J}_{ij}$ leads to
\BEA
{\overline{\det \chi^{-1}}}_{ij}&=&
\int\prod_{i=1}^N d\eta_i~d{\overline{\eta}}_i
\exp\left\{
-\frac{\alpha_p^2}{2 N}
\left(\sum_{i=1}^N{\overline{\eta}}_i\eta_i
\right)^2
+\sum_{i=1}^N\left(
\frac{1}{\beta(1-m_i^2)}+\alpha_p^2\beta(1-q)
\right)
\right\}
\\
&=&
\int\frac{dw}{\sqrt{2\pi/N}}\exp\left\{
N~H(w)
\right\}
\\
H(w)&=&-\frac{w^2}{2}+\frac{1}{N}
\sum_{i=1}^N\log\left(
iw\alpha_p
+\frac{1}{\beta(1-m_i^2)}+\alpha_p^2\beta(1-q)
\right)
\EEA
The saddle point of the above integral is determined by the following 
equations
\BEA
\frac{\p H}{\p w}&=&-w+\frac{1}{N}\sum_{i=1}^N
\frac{i\alpha_p}{iw\alpha_p
+\frac{1}{
\beta(1-m_i^2)}+\alpha_p^2\beta(1-q)}=0
\label{det1}
\\
\frac{\p^2 H}{\p w^2}&=&
-1+\frac{1}{N}\sum_{i=1}^N
\frac{\alpha_p^2}{
\left(iw\alpha_p+\frac{1}{\beta(1-m_i^2)}+\alpha_p^2\beta(1-q)\right)^2}
<0
\label{det2}
\EEA
Setting
$iv\alpha_p\equiv iw\alpha_p+\alpha_p^2\beta(1-q)
=iw\alpha_p+\frac{1}{\beta(1-m_i^2)}+\alpha_p^2\beta(1-q)$,
Eqs. (\ref{det1})-(\ref{det2}) read
\BEA
&&iv\left(
1-\frac{\alpha_p^2\beta^2}{N}\sum_{i=1}^N
\frac{(1-m_i^2)^2}{1+iv\alpha_p\beta(1-m_i^2)}
\right)=0
\label{v}
\\
&&1-\frac{\alpha_p^2\beta^2}{N}\sum_{i=1}^N\frac{(1-m_i^2)^2}
{1+iv\alpha_p\beta(1-m_i^2)}>0
\label{v2}
\EEA

Eq. (\ref{v}) has two solutions: 
$v=0$ and $v=v^\star$. In the first case the saddle point condition
Eq. (\ref{v2}) becomes the Plefka criterion Eq. (\ref{repli})
\BEQ
\nn
1-\frac{\alpha^2}{N}\sum_{i=1}^N\left(1-m_i^2\right)^2=x>0
\EEQ

If $v=v^\star$ Eq. (\ref{v2}) reduces to
\BEQ
iv\frac{\alpha_p^3\beta^2}{N}\sum_{i=1}^N
\frac{(1-m_i^2)^3}{(1+i v \alpha_p (1-m_i^2))^2}>0
\EEQ
Since $m_i\leq 1$, it must be $iv\in R^+$.
In order to make a  connection with the replica notation used in Refs.
[\onlinecite{BMan,Rieger}] we define 
$B=iv\alpha_p$ and Eq. (\ref{v}) becomes
\BEQ
1=\frac{\alpha_p^2\beta^2}{N}\sum_{i=1}^N
\frac{1-m_i^2}{1+B\beta(1-m_i^2)}<\frac{\alpha_p^2\beta^2}{N}
\sum_{i=1}^N(1-m_i^2)^2
\EEQ
Quite clearly the Plefka criterion is not satisfied for any
$B>0$. On the contrary setting $B=0$ the solution is consistent
with a physical solution.

\end{document}